\documentclass[twoside]{article}
\usepackage{fleqn,espcrc2}
\usepackage{amsmath}

\usepackage{graphicx}
\usepackage[figuresright]{rotating}

\title{The role of Monte Carlo within a diagonalization/Monte Carlo scheme}

\author{Dean Lee\address{Department of Physics, Univ. of Massachusetts, 
Amherst, MA 01003}}
       
\begin{document}

\begin{abstract}
We review the method of stochastic error correction which eliminates the truncation error associated with any subspace diagonalization. \ Monte Carlo sampling is used to compute the contribution of the remaining basis vectors not included in the initial diagonalization. \ The method is part of a new approach to computational quantum physics which combines both diagonalization and Monte Carlo techniques.
\vspace{1pc}
\end{abstract}

\maketitle

\section{INTRODUCTION}

In \cite{qse} and \cite{sec} a new approach was proposed for finding the
low-energy eigenstates of very large or infinite-dimensional quantum
Hamiltonians. \ This proposal combines both diagonalization and Monte Carlo
methods, each being used to solve a portion of the problem for which the
technique is most efficient. \ The first part of the proposal is to
diagonalize the Hamiltonian restricted to a subspace containing the most
important basis vectors for each low energy eigenstate. \ This may be
accomplished either through variational techniques or an \textit{ab initio}
method such as quasi-sparse eigenvector (QSE) diagonalization.\ \ The second
step is to include the contribution of the remaining basis vectors\ by Monte
Carlo sampling. \ The use of diagonalization allows one to consider systems
with fermion sign oscillations and extract information about wavefunctions and
excited states. \ The use of Monte Carlo provides tools to handle the
exponential increase in the number of basis states for large volume systems.

In this brief article we discuss the second half of the diagonalization/Monte
Carlo scheme. \ We discuss several new Monte Carlo techniques known as
stochastic error correction (SEC). \ There are two general varieties of
stochastic error correction, methods based on a series expansion and those
which are not. \ The series method starts with an eigenvector of the
Hamiltonian restricted to some starting subspace and then includes the
contribution of the remaining basis states as terms in an ordered expansion.
\ The idea is to form a perturbative expansion centered around a good
non-perturbative starting point.

We also discuss a technique called the stochastic Lanczos method. \ This
method again starts with eigenvectors of a Hamiltonian submatrix. \ Using
these as starting vectors, we define Krylov vectors, $\left|  j\right\rangle
,H\left|  j\right\rangle ,H^{2}\left|  j\right\rangle \cdots$, similar to
standard Lanczos diagonalization. \ The new ingredient is that matrix elements
between Krylov vectors, $\left\langle j^{\prime}\right|  H^{n}\left|
j\right\rangle ,$ are computed using matrix diffusion Monte Carlo. \ Since the
method does not rely on a series expansion, it has the advantage that the
starting vectors need not be close to the exact eigenvectors.

\section{SERIES METHOD}

Let $\left|  i\right\rangle $ be the eigenvectors of a Hamiltonian $H$
restricted to some subspace $S$. \ Let $\left|  A_{j}\right\rangle $ be the
remaining basis vectors in the full space not contained in $S$. \ We can
represent $H$ as
\begin{equation}
\left[
\begin{array}
[c]{ccccc}%
\lambda_{1} & 0 & \cdots & \left\langle 1\right|  H\left|  A_{1}\right\rangle
& \cdots\\
0 & \lambda_{2} & \cdots & \left\langle 2\right|  H\left|  A_{1}\right\rangle
& \cdots\\
\vdots & \vdots & \ddots & \vdots & \cdots\\
\left\langle A_{1}\right|  H\left|  1\right\rangle  & \left\langle
A_{1}\right|  H\left|  2\right\rangle  & \cdots &  E\cdot\lambda_{A_{1}} &
\cdots\\
\vdots & \vdots & \vdots & \vdots & \ddots
\end{array}
\right]  .
\end{equation}
We have used Dirac's bra-ket notation to represent the terms of the matrix.
\ In cases where the basis is non-orthogonal or the Hamiltonian is
non-Hermitian, the precise meaning of terms such as $\left\langle
A_{1}\right|  H\left|  1\right\rangle $ is the action of the dual vector to
$\left|  A_{1}\right\rangle $ upon the vector $H\left|  1\right\rangle $. \ We
have written the diagonal terms for the basis vectors $\left|  A_{j}%
\right\rangle $ with an explicit factor $E$ for reasons to be explained shortly.

Let us assume that $\left|  1\right\rangle $ is close to some exact
eigenvector of $H$ which we denote as $\left|  1_{\text{full}}\right\rangle $.
\ More concretely we assume that the components of $\left|  1_{\text{full}%
}\right\rangle $ outside $S$ are small enough so that we can expand in inverse
powers of the introduced parameter $E.$ \ In order to simply the expansion we
choose to shift the diagonal entries so that $\lambda_{1}=0.$

The series method of stochastic error correction is based on the $E^{-1}$
expansion,
\begin{equation}
\left|  1_{\text{full}}\right\rangle \propto\left[
\begin{array}
[c]{c}%
1\\
c_{2}^{\prime}E^{-1}+c_{2}^{\prime\prime}E^{-2}+\cdots\\
\vdots\\
c_{A_{1}}^{\prime}E^{-1}+c_{A_{1}}^{\prime\prime}E^{-2}+\cdots\\
\vdots
\end{array}
\right]  ,\label{unnorm}%
\end{equation}%
\begin{equation}
\lambda_{\text{full}}=\lambda_{1}^{\prime}E^{-1}+\lambda_{1}^{\prime\prime
}E^{-2}\cdots.
\end{equation}
It is convenient to choose the normalization of the eigenvector such that the
$\left|  1\right\rangle $ component remains 1. \ The convergence of the
expansion is controlled by the proximity of $\left|  1\right\rangle $ to
$\left|  1_{\text{full}}\right\rangle $. \ If $\left|  1\right\rangle $ is not
at all close to $\left|  1_{\text{full}}\right\rangle $ then it will be
necessary to use a non-series method such as the stochastic Lanczos method
discussed in the next section.  

The terms in the series are calculated by Monte Carlo sampling. \ All that is
required is an efficient way of generating random basis vectors $\left|
A_{k}\right\rangle $ with known probability rates. \ Let $P(A_{\text{trial}})$
denote the probability of selecting $\left|  A_{\text{trial}}\right\rangle $
on a given trial. \ If for example we are calculating the first order
correction to the eigenvalue, then we have%
\begin{align}
\lambda_{1}^{\prime} &  =-\sum_{j}\tfrac{\left\langle 1\right|  H\left|
A_{j}\right\rangle \left\langle A_{j}\right|  H\left|  1\right\rangle
}{\lambda_{A_{j}}}\\
&  =-\lim_{N\rightarrow\infty}\tfrac{1}{N}\sum_{i=1,\cdots,N}\tfrac
{\left\langle 1\right|  H\left|  A_{\text{trial}(i)}\right\rangle \left\langle
A_{\text{trial}(i)}\right|  H\left|  1\right\rangle }{\lambda_{A_{\text{trial}%
(i)}}P(A_{\text{trial}(i)})}.\nonumber
\end{align}

\section{STOCHASTIC LANCZOS}

We now consider a method called stochastic Lanczos which does not require the
starting vectors to be close to exact eigenvectors of $H.$ \ This is essential
if the eigenvectors of $H$ are not quasi-sparse and require extremely large
numbers of basis states to represent accurately.

Let $V$ be the full Hilbert space for our system. \ As in the previous section
let $S$ be the subspace over which we have diagonalized $H$ exactly. \ Let
$P_{S}$ be the projection operator for $S$ and let $\lambda_{j}$ and $\left|
j\right\rangle $ be the eigenvalues and eigenvectors of $H$ restricted to $S$
so that
\begin{equation}
P_{S}HP_{S}\left|  j\right\rangle =\lambda_{j}\left|  j\right\rangle .
\end{equation}
\ Let $Z$ be an auxiliary subspace, one which contains $S$ but excludes very
high-energy states. Let $P_{Z}$ be the projection operator for $Z$. \ We will
choose $Z$ such that $P_{Z}HP_{Z}$ is bounded above. \ Let$\ a$ be a real
constant which is greater than the midpoint of the minimum and maximum
eigenvalues of $P_{Z}HP_{Z}$. \ The stochastic Lanczos method uses the
operators $\left[  P_{Z}(H-a)P_{Z}\right]  ^{n}$ to approximate the low-energy
eigenvalues and eigenvectors of $P_{Z}HP_{Z}$. \ The goal is to diagonalize
$H$ in a subspace spanned by vectors%

\begin{equation}
\left|  d,j\right\rangle =\left[  P_{Z}(H-a)P_{Z}\right]  ^{d}\left|
j\right\rangle ,\label{slbasis}%
\end{equation}
for several values of $d$ and $j$. \ This requires calculating $\left\langle
d^{\prime},j^{\prime}\right|  \left.  d,j\right\rangle $ and $\left\langle
d^{\prime},j^{\prime}\right|  H\left|  d,j\right\rangle $. \ If our
Hamiltonian matrix is Hermitian, both of these terms can be written in the
general form
\begin{equation}
\left\langle j^{\prime}\right|  \left[  P_{Z}(H-a)P_{Z}\right]  ^{n}\left|
j\right\rangle .
\end{equation}
Therefore it suffices to determine the matrix
\begin{equation}
A_{n}\equiv P_{S}\left[  P_{Z}(H-a)P_{Z}\right]  ^{n}P_{S}.
\end{equation}
For non-orthogonal bases and non-Hermitian Hamiltonians, the only change is
that we use vectors
\begin{equation}
\left[  P_{Z}(H-a)P_{Z}\right]  ^{d}\left|  j\right\rangle
\end{equation}
to generate approximate right eigenvectors of $H$ and vectors in the dual
space
\begin{equation}
\left\langle j\right|  \left[  P_{Z}(H-a)P_{Z}\right]  ^{d}%
\end{equation}
to produce approximate left eigenvectors.  Adding and subtracting $P_{S}(H-a)P_{S}$, we can rewrite
\begin{equation}
A_{n}=P_{S}\left[
\begin{array}
[c]{c}%
P_{Z}(H-a)P_{Z}-P_{S}(H-a)P_{S}\\
+P_{S}(H-a)P_{S}%
\end{array}
\right]  ^{n}P_{S}.
\end{equation}

$A_{n}$ can now be evaluated recursively as
\begin{equation}
A_{n+1}=B_{n+1}+\sum_{m=0,\cdots,n}B_{m}(H-a)A_{n-m},
\end{equation}
where
\begin{equation}
B_{n}=P_{S}\left[  P_{Z}(H-a)P_{Z}-P_{S}(H-a)P_{S}\right]  ^{n}P_{S}.
\end{equation}

The components of $B_{n}$ are computed by matrix diffusion Monte Carlo. \ Any
general matrix product $M^{(1)}M^{(2)}\cdots M^{(n)}$ is a sum of degree $n$
monomials,
\begin{align}
& \left[  M^{(1)}M^{(2)}\cdots M^{(n)}\right]  _{jk}\label{sum}\\
& =\sum_{i_{1},\cdots i_{n-1}}M_{ji_{1}}^{(1)}M_{i_{1}i_{2}}^{(2)}\cdots
M_{i_{n-1}k}^{(n)}.\nonumber
\end{align}
We can interpret (\ref{sum}) as a sum over paths through the set of basis
vectors of $Z,$%
\begin{equation}
\left|  j\right\rangle \rightarrow\left|  i_{1}\right\rangle \rightarrow
\cdots\rightarrow\left|  i_{n-1}\right\rangle \rightarrow\left|
k\right\rangle \text{,}%
\end{equation}
with an associated weight $M_{ji_{1}}^{(1)}M_{i_{1}i_{2}}^{(2)}\cdots
M_{i_{n-1}k}^{(n)}$.

\section{Hubbard Model}

As an example of the new diagonalization/Monte Carlo method we consider the two-dimensional Hubbard model defined by
the Hamiltonian

\begin{align}
H  & =-t\sum_{<i,j>;\;\sigma=\uparrow,\downarrow}(c_{i\sigma}^{\dagger
}c_{j\sigma}+c_{j\sigma}^{\dagger}c_{i\sigma})\\
& +U\sum_{i}(c_{i\uparrow}^{\dagger}c_{i\uparrow}c_{i\downarrow}^{\dagger
}c_{i\downarrow}).\nonumber
\end{align}
The summation $<i,j>$ is over nearest neighbor pairs. $\ c_{i\sigma}^{\dagger
}$($c_{i\sigma}$) is the creation(annihilation) operator for a spin $\sigma$
electron at site $i$. $\ t$ is the hopping parameter, and $U$ controls the
on-site Coulomb repulsion. The model has attracted considerable attention in
recent years due to its possible connection to $d$-wave pairing and stripe
correlations in high-$T_{c}$ cuprate superconductors. \ In spite of its simple
form, the computational difficulties associated with finding the ground state
of the model are substantial even for small systems. \ Fermion sign problems
render Monte Carlo simulations ineffective for $U$ positive and away from
half-filling, and the collective effect of very large numbers of basis Fock
states make most diagonalization approaches very difficult. A brief overview
of the history and literature pertaining to numerical aspects of the Hubbard
model can be found in \cite{overview}.

As a test of our methods, we use QSE diagonalization with stochastic error
correction to find the ground state energy of the $4\times4$ Hubbard model
with $5$ electrons per spin. \ The corresponding Hilbert space has about
$2\cdot10^{7}$ dimensions. \ For the QSE diagonalization we use momentum Fock
states which diagonalize the quadratic part of the Hamiltonian. \ The
Hamiltonian is invariant under the symmetry group generated by reflections
about the $x$ and $y$ axes, interchanges between $x$ and $y,$ and interchanges
between $\downarrow$ and $\uparrow$. \ We find it convenient to work with
symmetrized Fock states. \ We will compute stochastic error corrections to
first order using the series method.

In Table 1 we present results for the ground state energy. \ We encountered no
trouble with the sign problem, and in fact one can easily see that each term
in the first order series expression is negative definite. \ The
energies are measured relative to the energy of the Fermi sea at $U=0$. \ The
errors reported are statistical errors associated with the first order SEC
calculation. \ Where available, we compare with the results presented in
\cite{huss}, which we label as Exact, Projector Quantum Monte-Carlo (PQMC),
and Stochastic Diagonalization (SD). \ Stochastic diagonalization is a
subspace diagonalization technique similar to QSE but one which uses a
different method for selecting the subspace and is based on a variational
principle \cite{deraedt}. \ Although the precise number of basis states used
in the SD calculations is not listed, we infer from numbers reported for a
modified $4\times4$ Hubbard system that roughly $10^{5}$ states were
used.\footnote{The discrete symmetries of the system were not utilized in
their calculations.}%

\begin{table}

\label{table:1}
\caption{Ground state energy of the $4\times4$ Hubbard model}
\begin{tabular}
[c]{|l|l|l|l|l|l|l|}\hline
Coupling & States & QSE & QSE+SEC & Exact & SD & PQMC\\\hline
$U=2t$ & $%
\begin{array}
[c]{c}%
100\\
500\\
1000
\end{array}
$ & $%
\begin{array}
[c]{c}%
-.4797\\
-.4945\\
-.5006
\end{array}
$ & $%
\begin{array}
[c]{c}%
-.50147(5)\\
-.50181(3)\\
-.50198(1)
\end{array}
$ & $-.50194$ & $-.5010$ & $-.44(5)$\\\hline
$U=4t$ & $%
\begin{array}
[c]{c}%
100\\
500\\
1000
\end{array}
$ & $%
\begin{array}
[c]{c}%
-1.620\\
-1.748\\
-1.800
\end{array}
$ & $%
\begin{array}
[c]{c}%
-1.8113(4)\\
-1.8242(3)\\
-1.8302(1)
\end{array}
$ & $-1.8309$ & $-1.829$ & $-1.8(2)$\\\hline
$U=5t$ & $%
\begin{array}
[c]{c}%
500\\
1000\\
2000
\end{array}
$ & $%
\begin{array}
[c]{c}%
-2.558\\
-2.651\\
-2.685
\end{array}
$ & $%
\begin{array}
[c]{c}%
-2.7073(4)\\
-2.7208(2)\\
-2.7231(1)
\end{array}
$ & $-2.7245$ & $-2.723$ & $-2.9(3)$\\\hline
\end{tabular}
\end{table}

Apparently QSE diagonalization with SEC handles the $4\times 4$ system quite well with relatively few states.  Much larger systems are being studied using both higher series corrections and stochastic Lanczos techniques \cite{salwen}.  Further examples of both stochastic error correction methods are presented in \cite{sec}.

\section*{ACKNOWLEDGMENTS}

The author thanks all collaborators on the works cited here and the organizers
and participants of the Lattice 2000 meeting in Bangalore. \ Financial support
provided by the National Science Foundation.

\end{document}